# A Reynolds-averaged turbulence modeling approach to the maintenance of the Venus superrotation

AKIRA YOSHIZAWA[1, 3a], HIROMICHI KOBAYASHI,[2, 3b] NORIHIKO SUGIMOTO,[2, 3c] NOBUMITSU YOKOI[4, 5d], and YUTAKA SHIMOMURA[2, 3e]

[1]3-2-10-306 Tutihasi, Miyamae-ku, Kawasaki 216-0005, Japan

[2]Department of Physics, Hiyoshi, Keio University, Yokohama 228-8521, Japan

[3]Research and Education Center for Natural Science, Hiyoshi, Keio University, Yokohama 228-8521, Japan

[4]Institute of Industrial Science, University of Tokyo, Komaba, Meguro-ku, Tokyo 153-8505, Japan

[5]Nordic Institute for Theoretical Physics (NORDITA), Roslagstullsbacken 23, 106 91 Stockholm, Sweden

[a]E-mail: ay-tsch@mbg.nifty.com
    Telephone: 81-44-854-7521    Fax: 81-44-854-7521
[b]E-mail: hkobayas@phys-h.keio.ac.jp
[c]E-mail: nori@phys-h.keio.ac.jp
[d]E-mail: nobyokoi@iis.u-tokyo.ac.jp
[e]E-mail: yutaka@phys-h.keio.ac.jp

A maintenance mechanism of an approximately linear velocity profile of the Venus zonal flow or superrotation is explored, with the aid of a Reynolds-averaged turbulence modeling approach. The basic framework is similar to that of Gierasch (1975) in the sense that the mechanism is examined under a given meridional circulation. The profile mimicking the observations of the flow is initially assumed, and its maintenance mechanism in the presence of turbulence effects is investigated from a viewpoint of the suppression of energy cascade. In the present work, the turbulent viscosity is regarded as an indicator of the intensity of the cascade. A novelty of this formalism is the use of the isotropic turbulent viscosity based on a nonlocal time scale linked to a large-scale flow structure. The mechanism is first discussed qualitatively. On the basis of these discussions, the two-dimensional numerical simulation of the proposed model is performed, with an initially assumed superrotation, and the fast zonal flow is shown to be maintained, compared with the turbulent viscosity lacking the nonlocal time scale. The relationship of the present model with the current general-circulation-model simulation is discussed in light of a crucial role of the vertical viscosity.

# 1. Introduction

Venus is quite similar to Earth in size and mass, but it possesses mysterious phenomena represented by the fast zonal flow, that is, the superrotation. Some prominent features of the Venus atmosphere are summarized here for later discussions. The dynamics and thermal structures of the atmosphere are discussed in the works by Crisp and Titov (1997), Gierasch (1997), Matsuda (2000), and Zasova et al. (2007). A comprehensive review of Venus is given by Bengtsson et al. (2013).

## 1.1. *Important characteristics linked to the superrotation*

The main constituent of the Venus atmosphere is $CO_2$ whose mass proportion is about 98 percent. The density is larger in the lower layer, and 50 and 90 percent of the total mass of $CO_2$ exist at the height lower than 10 km and 30 km, respectively. The surface is in a state of high temperature and pressure; namely, the temperature is about 735 K, and the pressure is about $9.2 \times 10^4$ hPa. The planetary rotation speed is about 1/200 of the Earth counterpart, and the temperature difference in the longitudinal direction is small at the lower atmosphere

The primary components of the Venus atmospheric motion are the zonal or longitudinal flow [figure 1 (Schubert 1983)] and the meridional circulation [figure 2 (Gierasch 1975)]. Some features of the former are summarized as follows (capital Z denotes zonal):

(Z1) The zonal flow is almost axisymmetric around the rotation axis. Its direction is from the east to west and is the same as that of the surface rotation.

(Z2) The flow is of about 100 ms$^{-1}$ at the upper part (height of about 65-70 km) of the clouds of sulfuric-acid aerosol, as is seen from figure 1. The Rossby number defined by the ratio of the velocity to the rotation speed at the equator is about 60, compared with about 0.1 of Earth.

(Z3) The zonal flow is observed down to the lower layer of height 5 km. It increases almost monotonically with height and is approximately subject to a solid-body-like rotation at each height.

Observations of the meridional circulation are scarce, compared with the zonal flow (Limaye 2007, Peralta et al. 2007, Sanchez-Lavega et al. 2008). The circulation is driven by the latitudinal temperature difference due to the solar heating and is thought to exist at the height up to 50 km (Grassi et al. 2007, Tellmann et al. 2008). Its velocity is about a few ms$^{-1}$ and is towards the poles in the upper layer.

**1.2.** *Current meteorological approaches to the superrotation*

**1.2.1. Theoretical approaches**. The generation and maintenance mechanisms of the superrotation are quite complicated, owing to a number of dynamical and thermal effects. It is difficult to identify the degree of importance of each effect in the observations. In these situations, any theoretical approach is inevitably subject to various simplifications and assumptions about the Venus atmospheric motion. In spite of such limitation, an advantage of a theoretical approach is the capability to extract specific factors and examine their relevance to the superrotation. A typical instance of those factors is the role of horizontal and vertical viscosities.

A representative theoretical approach to the superrotation is the work by Gierasch (1975). It is founded on three premises (capital G denotes Gierasch):

(G1) Flow quantities are axisymmetric around the rotation axis of Venus.

(G2) The meridional circulation is generated by the temperature difference between the equatorial and polar regions. It is towards the poles and equator at the upper and lower layers of the atmosphere, respectively; namely, a Hadley circulation occurs.

(G3) The horizontal viscosity is much larger than the vertical counterpart.

Under premise G-2, the meridional flow at the upper layer of the atmosphere transports the angular momentum towards the higher-latitude region, whereas the flow at the lower layer returns the momentum to the lower-latitude one. In the absence of a mechanism retarding the angular-momentum return, the momentum merely circulates through the meridional circulation. There is little room for its accumulation at the upper layer that gives rise to the superrotation.

In the context of premise G-3, the horizontal and vertical viscosities are denoted as $v_H$ and $v_V$, respectively. With $D$ as the scale height, the time scales linked to them are given by

$$\tau_H = \frac{D^2}{v_H}, \quad \tau_V = \frac{D^2}{v_V}, \tag{1}$$

in addition to the one-turn time scale of the meridional circulation, $\tau_M$. Among these three time scales, an inequality is assumed:

$$\tau_H \ll \tau_M \ll \tau_V. \tag{2}$$

Under the first inequality, parts of the angular momentum transported towards the high-latitude region by the meridional circulation is returned towards the low-latitude one before the circulation transports the momentum to the lower layer of the atmosphere. The second inequality guarantees that the vertical momentum transport is small in the one-circulation period. These processes pave the way for the occurrence of the superrotation that is a solid-body-like rotation at each height. A more detailed discussion about the time-scale inequality was made by Matsuda (1982).

**1.2.2. Numerical experiments**. In the mechanism by Gierasch (1975), a large horizontal viscosity plays an important role in the formation of a solid-body-like motion. Its theoretical origin, however, is not so clear there. A promising approach to the understanding of the superrotation under the least premises and assumptions is the numerical experiment based on the general circulation model (GCM) that has originally been developed in the study of the Earth atmospheric motion. In what follows, reference will be made to some recent numerical experiments of the superrotation. A comprehensive review of various current experiments is beyond the present scope and is left for the consultation of Bengtsson et al. (2013) and the works mentioned below.

An interesting instance of the GCM simulations is the work by Yamamoto and Takahashi (2003). Its primary premises are as follows (YT means Yamamoto and Takahashi):

(YT1) A zonally uniform heating is added below the layer at 80 km and is maximum at about 55 km.

(YT2) The vertical eddy viscosity $v_V$ is chosen as 0.15 m$^2$s$^{-1}$. The diffusion by the horizontal eddy viscosity $v_H$ is replaced with the fourth-order hyperviscosity. This replacement signifies that the vertical diffusion needs to be weaker than the horizontal one. This situation is consistent with inequality (2).

Yamamoto and Takahashi (2006, 2009) further examined effects of the three-dimensional solar heating with the intensity and maximum location different from those of Yamamoto and Takahashi (2003). The findings by Yamamoto and Takahashi (2003, 2006, 2009) may be summarized as follows:

(YTa) The maximum zonal flow with about 100 ms$^{-1}$ occurs near the top of clouds.

(YTb) A single-cell meridional circulation, that is, a Hadley circulation is formed, whose maximum speed is about 10 ms$^{-1}$.

(YTc) Fluctuations such as Rossby and gravity waves are generated, which are dominant at midlatitudes.

On the basis of these findings, the mechanism proposed by Gierasch may be interpreted as follows. The angular momentum at the low-height layer is pumped up by the meridional circulation. Parts of the momentum are returned towards the equator through the horizontal eddy-transport process due to the fluctuations such as Rossby and gravity waves. This process gives rise to the solid-body-like motion at each height and plays a role of a large horizontal viscosity in Gierasch (1975). Vertically propagating waves also transport the angular momentum towards the Venus surface, but its amount is much smaller than that by the meridional circulation.

In the foregoing GCM simulations, the meridional circulation plays a crucial role in the generation of the superrotation. A different maintenance mechanism was discussed by Takagi and Matsuda (2007) who used a nonlinear dynamical model. There the solar heating is applied at height about 70 km, and $\nu_V$ ranges from 0.0025 to 0.25 m$^2$s$^{-1}$. The fourth-order hyperviscosity is adopted for the horizontal diffusion. In this work, thermal tides are generated in the cloud layer and propagate towards the Venus surface. As a result, the angular momentum is transported downwards, and the zonal flow retrograde to the surface is induced. Through the surface friction, the angular momentum is pumped up to the atmosphere. Then the meridional circulation is not essential in the mechanism. The change of $\nu_V$ exerts little influence to the vertical structure of the superrotation. The increase in $\nu_V$, however, leads to weakening of the superrotation. This finding signifies that the weak vertical diffusion is indispensable for maintaining the superrotation of about 100 ms$^{-1}$.

In the GCM, an anisotropic turbulent viscosity is adopted, as was mentioned above. This is linked to the grid resolution. In the current GCM, mesh sizes are much larger in the horizontal direction, compared with the vertical one. As a result, horizontal small-scale disturbances have to be damped effectively for overcoming numerical instability. For instance, the operator $\nabla^n$ with an even integer $n$ ($n \geq 4$) is often applied to the horizontal diffusion-related terms. Such numerical formulations are based on experience, and their theoretical and observational foundations are not always so firm (Koshyk and Boer 1995). In a future situation with further developed computational capability, it is expected to be possible to simulate atmospheric phenomena under a more isotropic grid spacing. There a nearly isotropic viscosity or a new horizontal diffusion scheme might be incorporated (Vana et. al. 2008).

### 1.3. *Present approach: Reynolds-averaged turbulence modeling*

In the present work, a stance similar to Gierasch (1975) is taken. Namely, the following premises are adopted (P denotes present):

(P1) A single meridional circulation is assumed.

(P2) A zonal flow mimicking observations that monotonically increases with height is specified at the initial state of computation.

The observed zonal flow and meridional circuation occur through the complicated interaction between dynamical and thermal effects. Their generation processes are beyond the present scope, and our primary interest is to seek a maintenance mechanism of the approximately linear velocity profile of the zonal flow in the presence of diffusion effect intensified by turbulence. For the investigation into the mechanism, we shall adopt a Reynolds-averaged turbulence modeling approach based on an isotropic turbulent viscosity. This approach does not deny the foregoing findings by the GCM simulations about the relative magnitude of horizontal and vertical viscosities, as will be mentioned in subsection 5.4.

In the GCM simulations, large-scale motions of the Venus atmosphere contribute to the transport of angular momentum that leads to solid-body-like rotation. These motions tend to be quasi-two-dimensional, owing to planetary-rotation and stratification effects generating a large-scale balanced flow (thermal wind relation, vertical hydrostatic balance, horizontal cyclostrophic balance, etc.). The vertical spatial scale is much shorter than the horizontal one. Then the approach based on an isotropic turbulent viscosity is expected to be useful for investigating into the vertical diffusion arising from shorter scales in the superrotation.

In the present work, attention will be focused on the maintenance of a specified zonal-flow mimicking observations, in light of the suppression of the energy cascade that tends to make a flow structure uniform. The turbulent viscosity may be regarded as an indicator of the intensity of the cascade. This approach will be shown to shed light on the findings by the GCM simulations about the magnitude of the vertical viscosity.

The present work is organized as follows. In section 2, a system of fundamental equations is given. In section 3, the Reynolds-averaged turbulence modeling approach based on an isotropic turbulent viscosity is introduced, and the turbulent viscosity is discussed in light of the cascade of mean-flow energy. A time scale linked to a nonlocal flow structure is introduced, in terms of which the turbulent viscosity is modeled. In section 4, a maintenance mechanism of the superrotation is qualitatively discussed with the aid of the proposed model. In section 5, the model is solved numerically, and the results are discussed in the context of the GCM simulations. Concluding remarks are given in section 6. In appendix, reference is made to a method of synthesizing several time scales and constructing a comprehensive one.

## 2. *Fundamental equations*

A system of equations pertinent to the Venus atmospheric motion is given by

$$\frac{\partial \rho}{\partial t} + \nabla \cdot (\rho \mathbf{u}) = 0, \tag{3}$$

$$\frac{\partial}{\partial t} \rho u_i + \frac{\partial}{\partial x_j} \rho u_i u_j = \frac{\partial p}{\partial x_i} + \frac{\partial}{\partial x_j} \mu [s_{ij}]_{tr\ell} + \rho F_{Bi}, \tag{4}$$

where $\rho$ is the density, $\mathbf{u}$ is the velocity, $p$ is the pressure, $\mu$ is the molecular viscosity, $\mathbf{F}_B$ is the body force per unit mass, $s_{ij}$ is the velocity-strain tensor

$$s_{ij} = \frac{\partial u_j}{\partial x_i} + \frac{\partial u_i}{\partial x_j}, \tag{5}$$

and subscript trl denotes the traceless part defined by

$$[A_{ij}]_{tr\ell} = A_{ij} - \frac{1}{3} A_{\ell\ell} \delta_{ij}. \tag{6}$$

The right-hand side of equation (5) needs to be multiplied by a numerical factor 1/2 in the usual definition, which is dropped for the simplicity of resulting mathematical manipulation. Representative body forces are the Coriolis and buoyancy ones, which are written as

$$\mathbf{F}_B = 2\mathbf{u} \times \boldsymbol{\omega}_F + \mathbf{g}, \qquad (7)$$

where $\boldsymbol{\omega}_F$ is the angular velocity of the frame, and $\mathbf{g}$ is the gravitational acceleration.

The temperature linked to the buoyancy force obeys

$$\frac{\partial}{\partial t}\rho\theta_H + \nabla \cdot (\rho\theta_H \mathbf{u}) = -p\nabla \cdot \mathbf{u} + \nabla \cdot \left(\frac{\kappa}{C_P}\nabla\theta_H\right) + Q_H, \qquad (8)$$

where $\theta_H$ is the temperature, $C_P$ is the specific heat at constant pressure, $\kappa$ is the thermal conductivity, and $Q_H$ expresses various thermal effects such as the rate of solar heating etc. Equations (3), (4), and (8) constitute a closed system of equations through the addition of the thermodynamic relation for a perfect gas

$$p = (C_P - C_V)\rho\theta_H \qquad (9)$$

($C_V$ is the specific heat at constant volume).

In the present work, an established state of meridional circulation is assumed, as was noted in subsection 1.3, and attention is focused on a zonal flow whose velocity increases monotonically with height and is approximated by a linear profile. A primary driving force of the meridional circulation is the buoyancy force. In the present approach assuming the circulation, the force does not play an important role and is neglected. Moreover the Coriolis force is also dropped. This approach, however, does not deny its importance. It is highly probable that the force as well as the buoyancy one affects a generation mechanism of the meridional circulation. The circulation generated thus exerts influence to the zonal flow. In reality, these two forces are retained in the GCM simulation. With this point in mind, a detailed investigation is made upon effects of small-scale turbulence under a prescribed meridional circulation.

As was noted in subsection 1.1, about 90 percent of the total mass of $CO_2$ exists at the height lower than 30 km, and the Venus atmosphere is in a highly stratified state. The zonal flow tends to be more enhanced at the upper height at each latitude, owing to the conservation of the angular momentum carried by the meridional circulation. In this context, it is important to take the density stratification into account. In the present work, however, more attention is focused on a turbulent state of the atmospheric motion, and the density stratification is neglected. This formalism needs to be improved for a more quantitative discussion about the zonal flow, which is left for future work.

## 3. A Reynolds-averaged turbulence modeling approach based on an isotropic turbulent viscosity

### 3.1. *A simplified system of Reynolds-averaged equations*

The Reynolds averaging is applied in the simplified situation mentioned in section 2, where the density is assumed to be constant, and buoyancy and Coriolis forces are neglected. A quantity $f$ is divided into the mean $F$ and the fluctuation $f'$ as

$$f = F + f', \quad F = \langle f \rangle, \tag{10}$$

where

$$f = (\mathbf{u}, p, \boldsymbol{\omega}), \quad F = (\mathbf{U}, P, \boldsymbol{\Omega}), \quad f' = (\mathbf{u}', p', \boldsymbol{\omega}'), \tag{11}$$

($\boldsymbol{\omega}$ is the vorticity). The Reynolds averaging may be regarded as that around the rotation axis of Venus.

A Reynolds-averaged system of equations to be adopted is

$$\nabla \cdot \mathbf{U} = 0, \tag{12}$$

$$\frac{DU_i}{Dt} \equiv \left(\frac{\partial}{\partial t} + \mathbf{U} \cdot \nabla \right) U_i = -\frac{1}{\rho} \frac{\partial P}{\partial x_i} + \frac{\partial}{\partial x_j}\left(-R_{ij}\right) + \nu \nabla^2 U_i. \tag{13}$$

Here $R_{ij}$ is the Reynolds stress defined by

$$R_{ij} = \langle u'_i u'_j \rangle. \tag{14}$$

### 3.2. *A turbulent-viscosity approximation to the Reynolds stress*

The trace part of the Reynolds stress $R_{ij}$ is related to the turbulent kinetic energy

$$K = \frac{1}{2}\langle \mathbf{u}'^2 \rangle = \frac{1}{2} R_{ii}, \tag{15}$$

whereas the traceless counterpart is denoted as

$$B_{ij} = [R_{ij}]_{trl} = R_{ij} - \frac{2}{3} K \delta_{ij}. \tag{16}$$

The simplest model for $B_{ij}$ is the turbulent-viscosity representation based on an isotropic turbulent viscosity $\nu_T$, which is

$$B_{ij} = -\nu_T S_{ij}, \tag{17}$$

where $S_{ij}$ is the mean velocity-strain tensor given by

$$S_{ij} = \frac{\partial U_j}{\partial x_i} + \frac{\partial U_i}{\partial x_j}. \tag{18}$$

Within the theoretical framework based on the two-scale direct-interaction approximation (TSDIA) in wavenumber ($\mathbf{k}$) space (Yoshizawa 1984), $\nu_T$ is expressed as

$$\nu_T \propto \int_0^\infty dk \int_{-\infty}^s Q(k, \mathbf{x}; s, s', t) G(k, \mathbf{x}; s, s', t) ds' \quad (k = |\mathbf{k}|). \tag{19}$$

Here $Q$ and $G$ are the two-point velocity correlation function and the response or Green's one, respectively, and $s$ and $s'$ are the times associated with the correlation between fluctuations.

Equation (19) indicates that $\nu_T$ is composed of the contributions from fluctuations with various wavenumbers. In complicated turbulent flows, it is difficult to explicitly deal with these fluctuations in a theoretical manner, and $Q$ and $G$ need to be modeled in physical space. It is often approximated with the aid of $K$ and a turbulence time scale $\tau$ that characterize the intensity of velocity fluctuations and their characteristic time scales, respectively; namely, $\nu_T$ is written as

$$\nu_T = C_\nu K \tau, \qquad (20)$$

where $C_\nu$ is a model constant. A proper modeling of $\tau$ is crucial in the construction of the Reynolds stress based on the turbulent viscosity.

### 3.3. *Relationship of the turbulent viscosity with the cascade of mean-flow energy*

We consider the relationship of the turbulent viscosity $\nu_T$ with the enhancement of the cascade of mean-flow energy by turbulence effects. The turbulent kinetic energy $K$ [equation (15)] obeys

$$\frac{DK}{Dt} = P_K - \varepsilon + D_K . \qquad (21)$$

On the right-hand side, each term is the production, dissipation, and diffusion rates, respectively, which are defined by

$$P_K = -R_{ij} \frac{\partial U_j}{\partial x_i}, \qquad (22)$$

$$\varepsilon = \nu \left\langle \left( \frac{\partial u'_j}{\partial x_i} \right)^2 \right\rangle, \qquad (23)$$

$$D_K = \nabla \cdot \left( -\left\langle \left( \frac{1}{2} \mathbf{u}'^2 + \frac{p'}{\rho} \right) \mathbf{u}' \right\rangle \right) + \nu \nabla^2 K . \qquad (24)$$

The production rate $P_K$ is reduced to

$$P_K = \frac{1}{2} \nu_T S_{ij}^{\ 2}, \qquad (25)$$

under the turbulent-viscosity approximation; namely, $P_K$ is nonnegative and plays a role of sustaining a turbulent state against molecular viscous effects.

The relationship of $P_K$ with the energy cascade may be understood more clearly in light of the mean-flow energy $\mathbf{U}^2/2$. The latter is governed by

$$\frac{D}{Dt}\frac{\mathbf{U}^2}{2} = -P_K - \nu\left(\frac{\partial U_j}{\partial x_i}\right)^2 + \frac{\partial}{\partial x_i}\left(-\frac{P}{\rho}U_i - R_{ij}U_j\right) + \nu\nabla^2\frac{\mathbf{U}^2}{2}, \qquad (26)$$

from equation (13). The first two terms on the right-hand side extract the kinetic energy and transfer it to velocity fluctuations. The second term is much smaller at high Reynolds numbers. The action due to the first term corresponds to the energy cascade from the mean flow to the energy-containing velocity fluctuations in wavenumber space. Such energy cascade tends to destroy a structure of mean flow with a large velocity gradient or large $S_{ij}$ and make it uniform. Then $\nu_T$ is an indicator of the intensity of energy cascade.

The foregoing cascade process may be also seen from a viewpoint of the nonlinear interaction of $\mathbf{U}$. Equation (13) is reduced to

$$\frac{\partial U_i}{\partial t} = -\frac{\partial}{\partial x_i}\left(\frac{P}{\rho} + \frac{1}{2}\mathbf{U}^2\right) + (\mathbf{U}\times\mathbf{\Omega})_i + \frac{\partial}{\partial x_j}\left(-R_{ij}\right) + \nu\nabla^2 U_i. \qquad (27)$$

Then $\mathbf{U}\times\mathbf{\Omega}$ expresses the cascade of mean-flow energy through the nonlinear interaction. This point becomes clearer in the vorticity equation

$$\frac{\partial \Omega_i}{\partial t} = \nabla\times(\mathbf{U}\times\mathbf{\Omega})_i + \varepsilon_{i\ell m}\frac{\partial^2}{\partial x_j \partial x_\ell}\left(-R_{jm}\right) + \nu\nabla^2\Omega_i \qquad (28)$$

($\varepsilon_{ij\ell}$ is the alternating symbol). In the case of small $\mathbf{U}\times\mathbf{\Omega}$, the cascade of mean-flow energy is suppressed, and a nonuniform flow structure represented by $\mathbf{U}$ and $\mathbf{\Omega}$ tends to be maintained at high Reynolds numbers.

With the Venus atmospheric motion in mind, we introduce spherical coordinates $(r,\theta,\varphi)$, which denote radial, latitudinal, and longitudinal coordinates, respectively:

$$-\frac{\pi}{2} \leq \theta \leq \frac{\pi}{2}, \quad 0 \leq \varphi \leq 2\pi \qquad (29)$$

(the equatorial plane is denoted by $\theta = 0$). In this coordinate system, $\mathbf{U} \times \mathbf{\Omega}$ is given by

$$\mathbf{U} \times \mathbf{\Omega} = \left(U_\theta \Omega_\varphi - U_\varphi \Omega_\theta, U_\varphi \Omega_r - U_r \Omega_\varphi, U_r \Omega_\theta - U_\theta \Omega_r\right). \tag{30}$$

In the present work, attention is focused on $U_\varphi$ under a prescribed meridional circulation represented by $U_\theta$ and $U_r$. Then the maintenance of $U_\varphi$ characterizing the superrotation is closely associated with

$$(\mathbf{U} \times \mathbf{\Omega})_\varphi = U_r \Omega_\theta - U_\theta \Omega_r. \tag{31}$$

Equation (31) will be discussed in subsection 4.4 from this viewpoint.

It is interesting to consider the relationship of the cascade-related quantity $\mathbf{U} \times \mathbf{\Omega}$ with the helicity $\mathbf{U} \cdot \mathbf{\Omega}$. It is known well that $\mathbf{U} \cdot \mathbf{\Omega}$ is an indicator characterizing a helical flow structure. These two quantities are connected as

$$\frac{(\mathbf{U} \times \mathbf{\Omega})^2}{|\mathbf{U}|^2 |\mathbf{\Omega}|^2} + \frac{(\mathbf{U} \cdot \mathbf{\Omega})^2}{|\mathbf{U}|^2 |\mathbf{\Omega}|^2} = 1. \tag{32}$$

Equation (32) indicates that the magnitude of $\mathbf{U} \times \mathbf{\Omega}$ is affected by $\mathbf{U} \cdot \mathbf{\Omega}$. The cascade of mean-flow energy is suppressed in the region with large $\mathbf{U} \cdot \mathbf{\Omega}$, resulting in the maintenance of a nonuniform flow structure in a turbulent regime against diffusion effects.

The foregoing discussion shows that the helicity is a useful concept in the Reynolds-averaged approach. The mean-flow helicity $\mathbf{U} \cdot \mathbf{\Omega}$, however, is not a Galilean-invariant quantity and cannot be incorporated into the modeling in a straightforward manner. This situation is the reason why the concept of helicity has not been utilized in the approach. It has recently been shown (Yoshizawa et al. 2011) that the essence of the helicity may be taken into account, through the material derivative of $\mathbf{\Omega}$, that is, $D\mathbf{\Omega}/Dt$. One of the flows in which effects of helicity occur prominently is a swirling flow. In cylindrical coordinates $(x, r, \theta)$ with $x$ as the direction of swirling axis and the axisymmetry around it assumed, the radial component of $D\mathbf{\Omega}/Dt$ is approximated as

$$\left(\frac{D\mathbf{\Omega}}{Dt}\right)_r \cong -\frac{U_\theta \Omega_\theta}{r}. \tag{33}$$

In such a flow, $U_\theta \Omega_\theta$ is Galilean-invariant, but $U_x \Omega_x$ is not. The absence of $U_x$ in equation (33) signifies that $D\mathbf{\Omega}/Dt$ may capture the essence of the mean-flow helicity, without contradicting the Galilean-invariance requirement.

In the present work, $D\mathbf{\Omega}/Dt$ also plays a crucial role in a maintenance mechanism of the superrotation. There $D\mathbf{\Omega}/Dt$ is not linked to a helical flow property intrinsic to swirling motion, but the ability to capture such a coherent property is a primary motivation for its introduction into the investigation upon the superrotation. This point will be explained in subsection 4.2.

**3.4.** *A modeling of the turbulent viscosity with special attention to time scales*

In subsection 3.2, the importance of properly modeling a characteristic time scale $\tau$ was noted. In turbulent flows, various time scales occur. Of them, a familiar time scale associated with energy-containing velocity fluctuations is

$$\tau_E = \frac{K}{\varepsilon}. \tag{34}$$

This is the time scale during which kinetic energy $K$ is lost through the cascade. It has long been regarded as the primary one of turbulence. In the context of mean flow, two representative ones are

$$\tau_S = \frac{1}{\sqrt{S_{ij}^2}}, \tag{35}$$

$$\tau_\Omega = \frac{1}{\sqrt{\Omega_{ij}^2}}, \tag{36}$$

where the mean vorticity tensor $\Omega_{ij}$ is related to $\mathbf{\Omega}$ as

$$\Omega_{ij} = \frac{\partial U_j}{\partial x_i} - \frac{\partial U_i}{\partial x_j} = \varepsilon_{ij\ell}\Omega_\ell \quad \left(\Omega_{ij}^{\ 2} = 2\Omega^2\right). \tag{37}$$

Equations (35) and (36) characterize the straining and vortical motion of mean flow, respectively.

Here it should be stressed that $\mathbf{\Omega}$ ($\Omega_{ij}$) and $S_{ij}$ are the quantities characterizing local flow structures. A typical instance of nonlocal flow structures is a helical motion mentioned in subsection 3.3, where fluid streams while rotating. Nonlocal properties such as the degree of winding of a helical path line cannot be represented in terms of $\mathbf{\Omega}$ itself, and the usefulness of $D\mathbf{\Omega}/Dt$ was pointed out (Yoshizawa et al. 2011).

In the context of the superrotation discussed in section 4, the velocity increases almost monotonically over the layer whose thickness is of about 60 km. Such a nonlocal velocity profile may be characterized by both $U_\varphi$ (the longitudinal component of $\mathbf{U}$) and $\Omega_\theta$ (the latitudinal one of $\mathbf{\Omega}$), as will be shown by equations (58) and (62) below. Then the time scale dependent on both of them is necessary for the investigation into the superrotation.

With the ability of $D\mathbf{\Omega}/Dt$ to describe nonlocal flow structures in mind, we introduce the time scale

$$\tau_M = \frac{1}{\left\{(D\mathbf{\Omega}/Dt)^2\right\}^{1/4}} \tag{38}$$

(subscript M denotes material). It will be seen from equation (63) that $\tau_M$ is the time scale containing both $U_\varphi$ and $\Omega_\theta$.

The next important step in the turbulent-viscosity modeling is the construction of the single time scale $\tau$ that comprehends equations (34)-(36) and (38). A method for synthesizing them is not always unique. A simple and systematic one is explained in appendix. After the method, we have

$$\tau = \frac{K/\varepsilon}{\Lambda}. \tag{39}$$

Here the time-scale correction factor $\Lambda$ is given by

$$\Lambda = \sqrt{1 + C_S\left(\frac{K}{\varepsilon}S_{ij}\right)^2 + C_\Omega\left(\frac{K}{\varepsilon}\Omega_{ij}\right)^2 + C_{M\Omega}\left(\frac{K}{\varepsilon}\Omega\right)^2\left(\frac{K^2}{\varepsilon^2}\frac{D\Omega}{Dt}\right)^2}, \tag{40}$$

with positive model constants $C_S$ etc. given in equation (45). Equation (39) may be called the synthesized time scale.

The application of equation (39) to equation (20) results in

$$\nu_T = C_\nu \frac{K/\varepsilon}{\Lambda}, \tag{41}$$

that is,

$$B_{ij} = -C_\nu \frac{K^2/\varepsilon}{\Lambda} S_{ij}. \tag{42}$$

The ability of equation (41) to deal with nonlocal flow structures has already been demonstrated in a swirling pipe flow (Yoshizawa et al. 2011) and trailing vortices behind a wing tip (Yoshizawa et al. 2012)

### 3.5. *Summary of the proposed model*

The present model is composed of solenoidal condition (12), equation (13) for the mean velocity $\mathbf{U}$, and equations (16) and (42) for the Reynolds stress $R_{ij}$ with the time-scale correction factor $\Lambda$ [equation (40)].

Turbulence quantities $K$ and $\varepsilon$ occurring in $\nu_T$ are calculated from

$$\frac{DK}{Dt} = P_K - \varepsilon + \nabla \cdot \left\{\left(\nu + \frac{\nu_T}{\sigma_K}\right)\nabla K\right\}, \tag{43}$$

$$\frac{D\varepsilon}{Dt} = C_{\varepsilon 1}\frac{\varepsilon}{K}P_K - C_{\varepsilon 2}\frac{\varepsilon^2}{K} + \nabla \cdot \left\{\left(\nu + \frac{\nu_T}{\sigma_\varepsilon}\right)\nabla\varepsilon\right\}, \tag{44}$$

with $P_K$ given by equation (25). In equation (43), the last term arises from the modeling of equation (24). These turbulence equations are the same as those widely used in the current two-equation Reynolds-averaged turbulence modeling, except the difference of a mathematical expression for $\nu_T$ (Launder and Spalding 1974, Pope 2000).

The model constants are

$$C_\nu = 0.12, \quad C_S = 0.015, \quad C_\Omega = 0.02 C_S, \quad C_{M\Omega} = 0.30,$$

$$\sigma_K = 1.4, \quad C_{\varepsilon 1} = 1.5, \quad C_{\varepsilon 2} = 1.9, \quad \sigma_\varepsilon = 1.4, \tag{45}$$

which have been estimated from the application to constant-density turbulent flows.

## 4. Qualitative investigations into the maintenance of the superrotation

Qualitative discussions about the superrotation with the aid of observations and the proposed model will be instrumental to understanding the computed results presented in section 5.

### 4.1. *Flow quantities based on spherical coordinates*

In the spherical coordinates $(r,\theta,\varphi)$ in subsection 3.3, we introduce the height from the surface as

$$r = R_V + h. \tag{46}$$

Here $R_V \,(\cong 6000 \text{ km})$ is the radius of Venus, and $h$ is the height from the surface. Under the axisymmetry around the rotation axis, the steady mean flow $\mathbf{U}$ is written as

$$\mathbf{U} = \{U_r(r,\theta), U_\theta(r,\theta), U_\varphi(r,\theta)\}. \tag{47}$$

For simplifying the discussion about the superrotation, we consider the situation that

$$|U_\theta|, |U_\phi| \gg |U_r|. \tag{48}$$

This simplification is plausible except near the equatorial and polar regions (figure 2), and $U_\theta$ is approximately identified with the meridional circulation. Inequality (48), however, will be dropped in the numerical computation in section 5.

From equations (47) and (48), each component of $\mathbf{\Omega}$ is approximated as

$$\Omega_r = -\frac{1}{r\cos\theta}\frac{\partial}{\partial\theta}(U_\varphi \cos\theta) = -\frac{1}{r}\frac{\partial U_\varphi}{\partial\theta} + \frac{\tan\theta}{r}U_\varphi, \tag{49}$$

$$\Omega_\theta = -\frac{1}{r}\frac{\partial}{\partial r}rU_\varphi = -\frac{\partial U_\varphi}{\partial r} - \frac{U_\varphi}{r}, \tag{50}$$

$$\Omega_\varphi = \frac{1}{r}\frac{\partial}{\partial r}rU_\theta = \frac{\partial U_\theta}{\partial r} + \frac{U_\theta}{r}. \tag{51}$$

The material derivative $D\mathbf{\Omega}/Dt$ is similarly expressed in the form

$$\left(\frac{D\mathbf{\Omega}}{Dt}\right)_r = -\frac{U_\theta}{r}\frac{\partial \Omega_r}{\partial\theta} - \frac{U_\theta\Omega_\theta + U_\varphi\Omega_\varphi}{r}, \tag{52}$$

$$\left(\frac{D\mathbf{\Omega}}{Dt}\right)_\theta = -\frac{U_\theta}{r}\frac{\partial \Omega_\theta}{\partial\theta} + \frac{U_\theta\Omega_r}{r} - \frac{\tan\theta}{r}U_\varphi\Omega_\varphi, \tag{53}$$

$$\left(\frac{D\mathbf{\Omega}}{Dt}\right)_\varphi = -\frac{U_\theta}{r}\frac{\partial \Omega_\varphi}{\partial\theta} + \frac{U_\varphi\Omega_r}{r} + \frac{\tan\theta}{r}U_\varphi\Omega_\theta. \tag{54}$$

### 4.2. Order estimate of flow quantities

The order estimate of flow quantities is useful for qualitatively discussing the superrotation. Venus is covered by clouds mainly composed of sulfuric-acid aerosol, and available observational data are limited, as was referred to in subsection 1.1. In the following qualitative discussions based on the order estimate, details of observations are not always necessary.

The meridional circulation is approximated by $U_\theta$ under expression (48), which is

$$U_\theta = O(1) \text{ m s}^{-1} \quad (h < 50 \text{ km}). \tag{55}$$

The zonal flow $U_\varphi$ is observed in more detail. It increases almost monotonically with $h$ as

$$U_\varphi = 0 \sim 100 \text{ m s}^{-1} \quad (h = 0 \sim 60 \text{ km}), \tag{56}$$

as in figure 1 (Schubert 1983).

Equations (55) and (56) indicate that

$$\frac{U_\varphi}{U_\theta} = O(10) \sim O(10^2), \tag{57}$$

at the height where $U_\varphi$ becomes prominent. Namely, $U_\varphi$ is much higher than $U_\theta$ there. Then $\mathbf{U}$ is approximated by

$$\mathbf{U} \cong (0, 0, U_\varphi). \tag{58}$$

Equation (58) does not signify that the role of $U_\theta$ is minor in investigating into the dynamics of the superrotation. This point will be mentioned later.

From equation (56), we have

$$\frac{\partial U_\varphi}{\partial r} = \frac{\partial U_\varphi}{\partial h} = O(10^{-3}) \text{ s}^{-1} \quad (h = 0 \sim 60 \text{ km}), \tag{59}$$

$$\frac{U_\varphi}{r} \cong \frac{U_\varphi}{R_V} = O(10^{-5}) \text{ s}^{-1}, \tag{60}$$

which suggest

$$\frac{\partial U_\varphi / \partial r}{U_\varphi / r} \cong O(10^2). \tag{61}$$

In the comparison between $\partial f / \partial r$ and $(1/r)\partial f / \partial \theta$ ($f$ is a quantity such as $\mathbf{U}$), the latter is generally much smaller, owing to large $R_V$ in equation (46). In the following qualitative discussions, the term related to $\partial / \partial \theta$ are neglected. Equations (49)-(51) lead to

$$\mathbf{\Omega} \cong (0, \Omega_\theta, 0) \cong \left(0, -\frac{\partial U_\varphi}{\partial r}, 0\right), \tag{62}$$

from equations (57) and (61). In equations (52)-(54), the third term in equation (54) is specifically important from equations (58) and (62). Namely, $D\mathbf{\Omega}/Dt$ is approximated by

$$\frac{D\mathbf{\Omega}}{Dt} = \left(0, 0, \frac{\tan\theta}{r} U_\varphi \Omega_\theta\right), \tag{63}$$

which results in

$$\nu_T \cong C_\nu \frac{K^2/\varepsilon}{\sqrt{1 + C_S \left(\frac{K}{\varepsilon} S_{ij}\right)^2 + C_\Omega \left(\frac{K}{\varepsilon} \Omega_{ij}\right)^2 + C_{M\Omega} \left(\frac{K}{\varepsilon} \frac{\partial U_\varphi}{\partial r}\right)^2 \left(\frac{K^2}{\varepsilon^2} \frac{\tan\theta}{r} U_\varphi \frac{\partial U_\varphi}{\partial r}\right)^2}}, \tag{64}$$

from equations (40) and (41). Equation (63) clearly shows that $D\mathbf{\Omega}/Dt$ comprehends both $U_\varphi$ and $\Omega_\theta$ [the latter characterizes the monotonic increase in the former with height, as is seen from equation (62)]. This fact guarantees the statement in subsection 3.4 that the time scale $\tau_M$ [equation (38)] is a time scale characterizing the flow field of the superrotation.

### 4.3. *A maintenance mechanism of a fast zonal flow*

Let us investigate into the relationship of equation (64) with the maintenance of $U_\varphi$. With the observations of figure 1 in mind, the radial gradient $\partial U_\varphi/\partial r$ does not depend on $h$ so strongly on the whole and may be regarded as nearly constant. In equation (64), the ingredient suppressing $\nu_T$ is reduced to

$$\frac{\tan\theta}{r} U_\varphi \frac{\partial U_\varphi}{\partial r} = \frac{\tan\theta}{R_V + h} U_\varphi \frac{\partial U_\varphi}{\partial h} \propto U_\varphi, \tag{65}$$

which signifies that the degree of suppression of $\nu_T$ is governed by $U_\varphi$, except the equatorial and polar regions. Then $\nu_T$ is suppressed in the region with high $U_\varphi$, contributing to the maintenance of high $U_\varphi$. In the region with low $U_\varphi$, on the contrary, this suppression is weakened, and $U_\varphi$ is retarded.

The role of $U_\theta$ is not clear at the stage of the foregoing discussions. It is generated by the buoyancy force arising from the temperature difference between the equatorial and polar regions. The zonal flow $U_\varphi$ arises from the angular momentum carried by $U_\theta$. As was stated above, the suppression of $v_T$ is weakened at the low height, and the retardation of $U_\varphi$ occurs. Retarded $U_\varphi$ is supplied with a longitudinal momentum through $U_\theta$.

In summarizing these processes, a maintenance mechanism of the superrotation may be arranged in the following order (capitals M denote maintenance):

(M1) Existence of high and low $U_\varphi$ at the upper and lower layers of the atmosphere, respectively.

(M2) Suppression of $v_T$ at the upper layer and its weakening at the lower layer.

(M3) Maintenance of high $U_\varphi$ at the upper layer and retardation of $U_\varphi$ at the lower layer.

The superrotation is supposed to be maintained by the transport of angular momentum due to several disturbances in the equatorial region. In the GCM simulations mentioned in subsection in 1.2.2, they are identified with quasi-two-dimensional coherent vortices (barotropic and baroclinic eddies, and Rossby waves), thermal tides, gravity waves, etc. These processes are beyond the reach of the present work that focuses on the maintenance of the superrotation in a turbulent regime, under a prescribed meridional circulation.

### 4.4. *Relationship with the energy-cascade suppression*

From equations (58) and (62), equation (30) that governs the cascade of mean-flow energy in equation (27) is rewritten as

$$\mathbf{U} \times \mathbf{\Omega} = \left(-U_\varphi \Omega_\theta, 0, 0\right) = \left(\frac{\partial}{\partial r} \frac{U_\varphi^2}{2}, 0, 0\right). \tag{66}$$

Then $\mathbf{U} \times \mathbf{\Omega}$ does not possess the $\varphi$ component [equation (31)] that exerts influence to $U_\varphi$.

With equation (48) about the relative magnitude of velocity components in mind, the $r$ component of equation (27) may be written approximately as

$$-\frac{\partial}{\partial r}\left\{P + \frac{1}{2}\left(U_\theta^2 + U_\varphi^2\right)\right\} + \frac{\partial}{\partial r}\frac{1}{2}U_\varphi^2 = 0, \qquad (67)$$

which leads to

$$P + \frac{1}{2}U_\theta^2 = \Gamma(\theta), \qquad (68)$$

where $\Gamma$ is an arbitrary function of $\theta$. In short, $\mathbf{U} \times \mathbf{\Omega}$ is absorbed into the pressure balance equation in the radial direction, and its role of cascading energy is lost. The suppression of the cascade corresponds to that of $\nu_T$ through $D\mathbf{\Omega}/Dt$.

## 5. Quantitative discussions based on the computation of the model

The qualitative discussions of section 4 based on inequality (48) may not give a definite information about $\nu_T$ that is associated with the GCM simulation of the superrotation. For obtaining a quantitative insight into the relationship with the GCM simulation, we need to perform the numerical computation of the present model. In accordance to the premises of the qualitative discussions in subsection 4.3, a meridional circulation and a zonal flow mimicking observations are specified, and attention is focused on the maintenance of the latter. Under a specified meridional circulation, Iga and Matsuda (1999) performed a two-dimensional numerical simulation and pointed out that the superrotation may be maintained by the upward transport of angular momentum by the meridional circulation.

In the following computation, equation (13) for $U_\varphi$ is computed in the combination with equations (43) and (44) for $K$ and $\varepsilon$, where inequality (48) is dropped.

## 5.1. *Specification of the meridional circulation*

Two velocity components of the meridional circulation are expressed in terms of the stream function $\psi$ as

$$U_\theta = \frac{1}{r\cos\theta}\frac{\partial \psi}{\partial r}, \tag{69}$$

$$U_r = -\frac{1}{r^2 \cos\theta}\frac{\partial \psi}{\partial \theta}, \tag{70}$$

in the coordinates with the equatorial pane as $\theta = 0$.

For $\psi$, we adopt

$$\psi = U_M R_C{}^2 \left(\frac{r - R_V}{R_C}\right)^2 \frac{r - R_C}{R_C} \cos\theta \sin 2\theta. \tag{71}$$

Here $U_M$ is the reference velocity characterizing the magnitude of meridional circulation, and $R_C$ is the distance of the upper part of the clouds from the center of Venus. They are chosen as

$$U_M = 60600 \text{ m s}^{-1}, \quad R_C = (R_V + 60) \text{ km}. \tag{72}$$

The reference velocity $U_M$ corresponds to be about $U_\theta = 6$ ms$^{-1}$ at $r = R_C$ and $\theta = 45°$. Figure 3 shows the contour lines of $\psi$ given by equation (71).

From equations (69)-(71), we have

$$U_\theta = U_M \frac{R_C}{r} \frac{r - R_V}{R_C} \frac{3r - R_V - 2R_C}{R_C} \sin 2\theta, \tag{73}$$

$$U_r = 2U_M \left(\frac{R_C}{r}\right)^2 \left(\frac{r - R_V}{R_C}\right)^2 \frac{r - R_C}{R_C}\left(3\sin^2\theta - 1\right). \tag{74}$$

At $\theta = (10°, 45°, 80°)$, their profiles via the height $h\,(= r - R_V)$ are shown in figure 4. At $\theta = 45°$, $U_\theta$ is much larger than $U_r$ and may be regarded as the meridional-circulation velocity.

### 5.2. Boundary conditions on the zonal flow and turbulence quantities

The computation is made in a time-marching manner. As initial $U_\varphi$, a linear profile

$$U_\varphi = \left(U_Z - U_\varphi|_{r=R_V}\right)\frac{r - R_V}{R_C - R_V} + U_\varphi|_{r=R_V} \quad \left(U_Z = 100 \text{ m s}^{-1}\right) \tag{75}$$

is chosen in light of the observations shown in figure 1. Equation (75) is shown in figure 5, where $U_\varphi$ is assumed to be $100 \text{ m s}^{-1}$ at $h = 60$ km (the upper part of the clouds). It is constant at each height. Our primary concern is whether this profile is really maintained in a turbulent regime or not.

The boundary conditions are given as follows (capital BC denotes boundary condition):

(BC1) At the surface, $U_\varphi$ is equal to the velocity of Venus, that is,

$$U_\varphi|_{r=R_V} = \Omega_V R_V \cos\theta, \tag{76}$$

where $\Omega_V$ is the angular velocity of Venus whose rotational period is 243 days. For $K$ and $\varepsilon$, we take

$$\frac{\partial K}{\partial r} = 0, \quad \varepsilon = P_K \quad \left(P_K = -R_{ij}\frac{\partial U_j}{\partial x_i}\right). \tag{77}$$

(BC2) At the upper part of the clouds, the free-slip conditions

$$\frac{\partial U_\varphi}{\partial r} = \frac{\partial K}{\partial r} = \frac{\partial \varepsilon}{\partial r} = 0 \tag{78}$$

are imposed.

We are in a position to refer to condition (77). Strictly speaking, $K$ at the surface needs to obey the noslip condition. In the analysis of engineering flows, Reynolds-averaged models of the present type are supplemented with molecular-viscosity effects, and the condition is fulfilled. In meteorological phenomena whose spatial scales are huge, the explicit treatment of those effects is not realistic. Equation (77) is an approximate method of incorporating the turbulent-production process near a solid boundary, without explicitly dealing with its close vicinity. The second relation comes from equation (43) for $K$ through the discard of the advection and diffusion terms. Near a solid boundary, they are less important than the production and dissipation terms.

In the computation of engineering turbulent flows, the $K-\varepsilon$ model is frequently used, which is obtained from equation (41) through the use of $\Lambda = 1$ and $C_\nu = 0.09$ (Launder and Spalding 1974, Pope 2000). The $K-\varepsilon$ model is similar to a model in the hierarchy by Mellor and Yamada (1974) that is utilized in the computation of atmospheric boundary layers. The comparison between these two results will elucidate the essence of the present model, specifically, the role of the nonlocal time scale $\tau_M$ [equation (38)].

### 5.3. *Numerics*

The number of grid points in the $r$ and $\theta$ directions is set to $(N_r, N_\theta) = (64, 64)$. It is confirmed that the results shown below do not change for $(N_r, N_\theta) = (512, 512)$. For the spatial discretization, the second-order central-finite-difference method and the staggered grid system are adopted. There $U_r$ and $U_\theta$ are defined at the cell side, whereas $U_\varphi$, $K$, and $\varepsilon$ are at the cell center.

The time marching scheme is the first-order Euler method with the time step $\Delta t = 10^{-3}$ s for the $K-\varepsilon$ model and $\Delta t = 10^{-1}$ s for the present and constant-viscosity ones. The temporal integration is made up to $10^7 \Delta t$ for the $K-\varepsilon$ model and $10^6 \Delta t$ for the other two. It is not always long enough in the sense of obtaining a steady-state solution from an arbitrary initial condition. Then a definite conclusion cannot be drawn about the steady state. In the comparison with the $K-\varepsilon$ model, however, the computed results will show the significance of $D\mathbf{\Omega}/Dt$ effects on $\nu_T$ in a maintenance mechanism of the superrotation.

Graphic Processing Unit, Nvidia Tesla C2050 (GPU) is used in the computation. It implements 60 times faster than a CPU (Intel Xeon E5607, 2.27GHz) where the code is optimized for a single CPU.

### 5.4. *Computed results*

Figure 6 gives the zonal velocity $U_\varphi$ at $\theta = (10°, 45°, 80°)$ computed by the present model. A small amount of change is observed at $\theta = (10°, 80°)$, but the initial profile is nearly maintained. In the height ranging from 30 to 40 km, $U_\varphi$ increases towards the high latitude because of the meridional circulation. The maintenance of the initial profile is due to $D\boldsymbol{\Omega}/Dt$ in the nonlocal time scale $\tau_M$ [equation (38)].

In order to clarify the foregoing situation, we consider the $K-\varepsilon$ model that is frequently used in the computation of engineering turbulent flows and is similar to a model in the hierarchy by Mellor and Yamada (1974). The model predicts uniform $U_\varphi$, as is in figure 7. The difference between these two computed results is attributed to $K$ and the turbulence time scale $\tau$ [equation (38)] that constitutes $\nu_T$ [equation (20)]. The selection of $\tau$ is specifically important. Figure 8 shows $K$ at $\theta = 45°$ computed by the present and $K-\varepsilon$ models. There $K$ by the $K-\varepsilon$ model is about $10^4$ times larger than that of the present one. Moreover the turbulence time scale predicted by the $K-\varepsilon$ model, that is, equation (39) with $\Lambda=1$, is about $10^3$ times larger than that of the present model. As a result, the magnitude of $\nu_T$ by the $K-\varepsilon$ model is $10^7$ times the present counterpart, as is seen from figure 9. Under such an intense turbulent diffusion effect, the initial profile of $U_\varphi$ disappears rapidly and becomes uniform, except a thin layer in the close vicinity of the surface (the layer may not be shown in the scale of figure 7). This thin layer resembles the surface layer of a planetary boundary layer, where turbulent diffusion effects are strong. The thickness of the surface layer is 50-100 m, which is very thin, compared with that of the whole layer (400-1000 m). Through the surface layer, the mean flow steeply changes into the outer-layer counterpart, resulting in a gradually-varying velocity profile in the latter. From these findings, it may be concluded that small $\nu_T$, specifically, $\nu_T$ decreasing rapidly with height is crucial in maintaining an approximately linear profile of the superrotation in a turbulent regime.

The turbulent viscosity $v_T$ at $\theta = (10°, 45°, 80°)$ assessed by the present model is shown in figure 10. It has a spatial distribution in the latitudinal direction, becoming small in the region away from the equator. In the numerical studies using the GCM, a constant turbulent viscosity has been used for the simplicity of computation.

Let us see the influence of a constant turbulent viscosity on $U_\varphi$. There equations (43) and (44) for $K$ and $\varepsilon$ are discarded. Figure 11 shows $U_\varphi$ at $\theta = (10°, 45°, 80°)$ computed for two cases of the horizontal viscosity $v_H$ and the vertical one $v_V$: (a) $v_H = v_V = 10^3$ m$^2$s$^{-1}$ and (b) $v_H = 10^3$ m$^2$s$^{-1}$ and $v_V = 0.15$ m$^2$s$^{-1}$. Case (a) shows that high $v_V$ leads to the retardation of $U_\varphi$ in the height from 40 to 60 km. In Case (b), on the other hand, low $v_V$ gives the profiles quite similar to that by the present model. This tendency is consistent with Takagi and Matsuda (2007) who used constant $v_V$ ranging from 0.0025 to 0.25 m$^2$s$^{-1}$ in the nonlinear dynamical model. They pointed out that the increase in $v_V$ weakens the superrotation. In the present model, the magnitude of spatially varying $v_V$ changes from $10^{-4}$ to 0.1 in the latitude direction. It is interesting that the predicted magnitude is consistent with the work by Takagi and Matsuda (2007).

In the GCM simulation by Yamamoto and Takahashi (2003), $v_H$ is replaced with the hyperviscosity in addition to the choice of $v_V = 0.15$ m$^2$s$^{-1}$. The hyperviscosity is indispensable for the computation based on large mesh sizes in the horizontal plane. In the present work, an isotropic turbulent viscosity, that is, $v_H = v_V$, is adopted, and $v_T$ consistent with $v_V$ in the GCM simulations is obtained. Then the incorporation of present $v_V$ into the horizontal viscosity in the GCM does not affect the hyperviscosity parts since the former is much smaller. In the GCM simulation where horizontal and vertical mesh sizes become comparable, $v_H$ may be expected to be replaced with the viscosity of the present type.

The computation time of the present model is short, compared with that in the GCM simulation. Then we cannot draw a definite conclusion about a steady state of solution. Under this constraint, however, an important feature of the model was clarified through the foregoing comparison with the $K-\varepsilon$ model.

In order to further show the usefulness of the model, we discuss effects of the magnitude of meridional circulation. We adopt the meridional velocity that is ten times the original one (figure 4) and is comparable to the zonal flow. The computed results based on such an unrealistically fast meridional circulation is given in figure 12. It shows that the initially-assumed linear velocity profile of $U_\varphi$ is lost in the present short computation time. The change from the initial profile is prominent at middle and high latitude. Specifically, $U_\varphi$ is almost uniform at the height between 30 km and 50 km. Its cause may be seen from figure 13, which gives large $\nu_T$ at the height. Namely, the strong diffusion due to large $\nu_T$ erases the linear profile intrinsic to the superrotation, making it uniform.

The foregoing finding may be explained from a viewpoint of the cascade of mean-flow energy. The nonlinear interaction associated with the cascade is given by $(\mathbf{U}\times\mathbf{\Omega})_\varphi$ [equation (31)]. In the simplified situation in subsection 4.2, it is reduced to

$$U_r \Omega_\theta - U_\theta \Omega_r \cong -\frac{\tan\theta}{r} U_\theta U_\varphi, \tag{79}$$

where the second term dependent on $U_\varphi$ itself is retained in equation (49). A noteworthy feature of equation (79) is the dependence on $U_\theta$ and $\tan\theta$. The use of an unrealistically fast meridional flow leads to an increase in the nonlinear interaction linked to the energy cascade. The dependence on $\tan\theta$ signifies that the increase is more prominent at a higher latitude. These facts are consistent with the computed results in figures 12 and 13.

In summary, the computation with an unrealistically fast meridional flow indicates that the meridional circulation with $O(1)\,\mathrm{ms}^{-1}$ is consistent with the observations of the zonal flow.

## 6. Concluding remarks

In the present work, we investigated into an aspect of the superrotation, that is, the maintenance of a fast zonal flow in a turbulent regime. There a meridional circulation was specified, and the temporal variation of an initial zonal-flow profile mimicking observations was examined. A maintenance mechanism of the zonal flow was sought with resort to a Reynolds-averaged turbulence modeling approach. In the approach, special attention was paid to the fact that the observed velocity of the superrotation increases almost monotonically over the whole layer. A nonlocal time scale appropriate for describing such a flow structure was introduced, in terms of which the turbulent viscosity is modeled.

On the basis of the proposed model, we first discussed the maintenance mechanism in a qualitative manner. The mechanism was attributed to the suppression of the turbulent viscosity and the energy cascade.

Next we calculated the model numerically and confirmed that an initial zonal-flow profile is really sustained. Specifically, the magnitude of the calculated turbulent viscosity was examined, which shows that it is consistent with the vertical viscosity in the current GCM simulations. This finding is considered to give a theoretical support to the latter and a suggestion to the choice of the viscosity in the GCM. Moreover the computation based on an unrealistically fast meridional circulation supports the coexistence of the observed superrotation and meridional circulation.

From these discussions, it may be concluded that the present Reynolds-averaged turbulence modeling approach is a useful tool for bridging the theoretical approach such as Gierasch (1975) and the numerical simulation by the GCM.

**Acknowledgments**

The authors are grateful to the referees for improving the presentation of the article.**References**

Bengtsson, L., Bonnet, R-M. Grinspoon, D., Koumoutsaris, S., Lebonnois, S., and Titov, D. V. (Editors), *Towards Understanding the Climate of Venus: Applications of Terrestrial Models to Our Sister Planet*, 2013 (Springer: New York).

Crisp, D. and Titov, D., The thermal balance of the Venus atmosphere. In V*enus II: Geology, Geophysics, Atmosphere, and Solar Wind Environment*, edited by S. W. Bougher, D. M. Hunten and R. J. Phillips, pp. 353-384, 1997 (University of Arizona Press: Tucson).

Gierasch, P. J., Meridional circulation and maintenance of the Venus atmospheric rotation. *J. Atmos. Sci.* 1975, **32**, 1038-1044.

Gierasch, P. J., Goody, P. M., Young, R. E., Crisp, D. et al., The general circulation of the Venus atmosphere: An assessment. In V*enus II: Geology, Geophysics, Atmosphere, and Solar Wind Environment*, edited by S. W. Bougher, D. M. Hunten and R. J. Phillips, pp. 459-500, 1997 (University of Arizona Press: Tucson).

Grassi, D., Drossart, P., Piccioni, G., Ignatiev, N. I., et al., Retrieval of air temperature profiles in the Venus atmosphere from VIRTIS-M data: Description and validation of algorithms. *J. Geophys. Res.* 2008, **113**, E00B09.

Iga, S. and Matsuda, Y., A mechanism of the super-rotation in the Venus atmosphere: Meridional circulation and barotropic instability. *Theor. Appl. Mech.* 1999, **48**, 379-383.

Koshyk, J. N. and Boer, G. J., Parameterization of dynamical subgrid-scale processes in a spectral GCM. *J. Atoms. Sci.* 1995, **52**, 965-976.32

Zasova, L. V., Ignatiev, N. I., Khatuntsev, I.A., and Linkin, V., Structure of the Venus atmosphere. *Planet. Space Sci.* 2007, **55**, 1712-1728.

**Appendix: A method of time-scale synthesis**

As is seen from the discussion about inequality (2), a shorter time scale is more important, in general, for characterizing a flow structure. We consider the condition under which nonlocal time scale $\tau_M$ [equation (38)] becomes more important than the local one $\tau_\Omega$ [equation (36)] or vice versa. For evaluating the relative importance of $\tau_M$ and $\tau_\Omega$, we adopt $\tau_E$ [equation (34)] as a reference one and introduce

$$\tau_{\Omega M} = \tau_\Omega \left( \frac{\tau_M}{\tau_E} \right)^2. \tag{A.1}$$

In case that $\mathbf{\Omega}$ varies slowly along the mean-flow path and that $\tau_M$ is longer than $\tau_E$, the importance of equation (A.1) is lost, compared with $\tau_\Omega$. Namely, the nonlocal vortical property of a flow becomes less important than the local one. Exponent 2 in equation (A.1) is chosen from the requirement that the expression resulting from the following time-scale synthesis is of an analytic form in $D\mathbf{\Omega}/Dt$ [see equation (40)].

An important step in the turbulent-viscosity modeling is the construction of the time scale $\tau$ necessary in modeling $\nu_T$ [equation (20)]. A method for synthesizing several scales was previously discussed with the aid of the TSDIA (Yoshizawa *et al.* 2006). After the method, we adopt

$$\frac{1}{\tau^2} = \frac{1}{\tau_E^2} + C_S \frac{1}{\tau_S^2} + C_\Omega \frac{1}{\tau_\Omega^2} + \frac{C_{M\Omega}}{2} \frac{1}{\tau_{M\Omega}^2}, \tag{A.2}$$

with positive constants $C_S$ etc., leading to equations (39) and (40). It should be stressed that a method of synthesizing time scales is not unique.

In equation (A.2), the exponents of $\tau$ etc. were chosen as 2 like $\tau^2$. This choice is required from the matching with the theoretical finding by the TSDIA that is formulated in the weak-inhomogeneity limit. The state of weak inhomogeneity is defined by

$$\frac{\tau_E}{\tau_S} = \sqrt{\left(\frac{K}{\varepsilon}S_{ij}\right)^2} \ll 1, \quad \frac{\tau_E}{\tau_\Omega} = \sqrt{\left(\frac{K}{\varepsilon}\Omega_{ij}\right)^2} \ll 1,$$

$$\frac{\tau_E}{\tau_{M\Omega}} = \left\{\frac{K^4}{\varepsilon^4}\left(\frac{D\mathbf{\Omega}}{Dt}\right)^2\right\}^{1/4} \ll 1. \tag{A.3}$$

In this limit, equation (41) with equation (40) as $\Lambda$ may be expanded as

$$\nu_T = C_\nu \frac{K^2}{\varepsilon}\left[1 - \frac{C_S}{2}\left(\frac{K}{\varepsilon}S_{ij}\right)^2 - \frac{C_\Omega}{2}\left(\frac{K}{\varepsilon}\Omega_{ij}\right)^2 - \frac{C_{M\Omega}}{2}\left(\frac{K}{\varepsilon}\mathbf{\Omega}\right)^2\left(\frac{K^2}{\varepsilon^2}\frac{D\mathbf{\Omega}}{Dt}\right)^2\right.$$

$$\left. + \frac{3}{8}\left\{C_S\left(\frac{K}{\varepsilon}S_{ij}\right)^2 + C_\Omega\left(\frac{K}{\varepsilon}\Omega_{ij}\right)^2\right\}^2 + \cdots\right]. \tag{A.4}$$

Here attention is focused on the first three parts in the parenthesis. They are coincident with the terms that were previously derived from the TSDIA with the Kolmogorov spectrum imbedded (Yoshizawa 1984, Okamoto 1994). The choice of exponent 2 such as $\tau^2$ in equation (A.2) is justified from this matching, and other exponents are ruled out.

**Figure captions**

Figure 1. Observations of Venus zonal flow (Schubert 1983).

Figure 2. Meridional circulation (solid line) and zonal flow (broken line) (Gierasch 1975).

Figure 3. Stream function $\psi$ [equation (63)] with the contour level of $5 \times 10^5$ m$^3$s$^{-1}$ ($\psi = 0$ at $r = R_C$).

Figure 4. Meridional circulation at $\theta = (10°, 45°, 80°)$ : equations (65) and (66).

Figure 5. Initial profile of the zonal flow given at $\theta = (10°, 45°, 80°)$ : equation (67)

Figure 6. Zonal-flow profiles at $\theta = (10°, 45°, 80°)$: present model.

Figure 7. Zonal-flow profiles at $\theta = (10°, 45°, 80°)$: $K - \varepsilon$ model.

Figure 8. Turbulent kinetic energy at $\theta = 45°$: present and $K - \varepsilon$ models.

Figure 9. Turbulent viscosity at $\theta = 45°$: present and $K - \varepsilon$ models.

Figure 10. Turbulent viscosity at $\theta = (10°, 45°, 80°)$: present model.

Figure 11. Zonal flows by constant-viscosity models at $\theta = (10°, 45°, 80°)$ : $\nu_H = \nu_V = 10^3$ m$^2$s$^{-1}$ (left); $\nu_H = 10^3$ m$^2$s$^{-1}$, $\nu_V = 0.15$ m$^2$s$^{-1}$ (right)

Figure 12. Zonal-flow profiles at $\theta = (10°, 45°, 80°)$ with ten times the original meridional circulation: present model.

Figure 13. Turbulent viscosity at $\theta = (10°, 45°, 80°)$ with ten times the original meridional circulation: present model.

Fig. 1

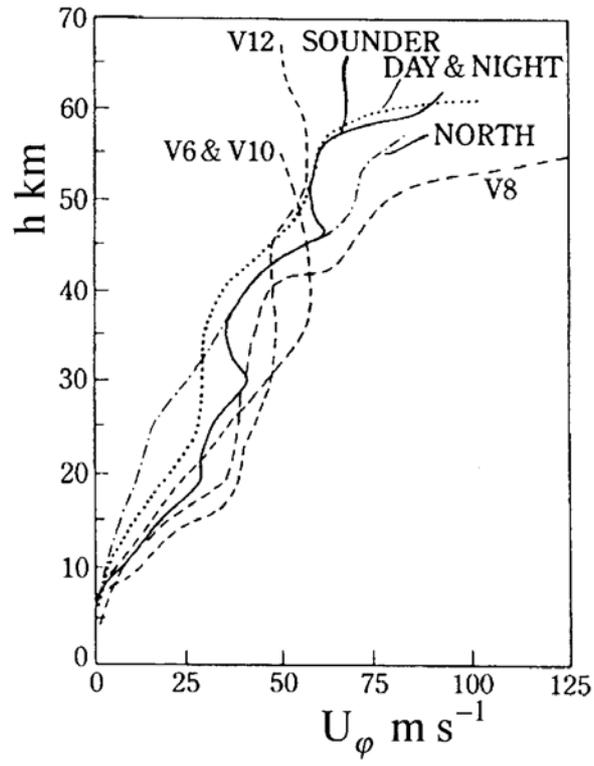

Fig. 2

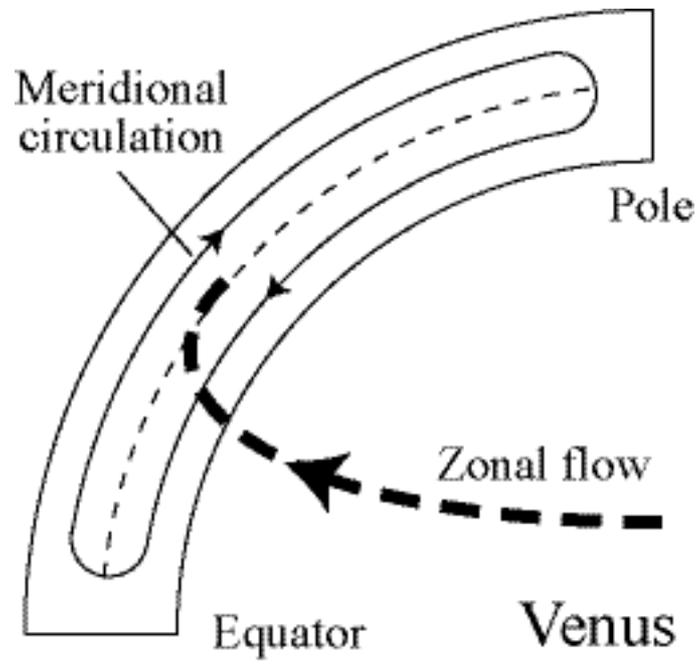

Fig. 3

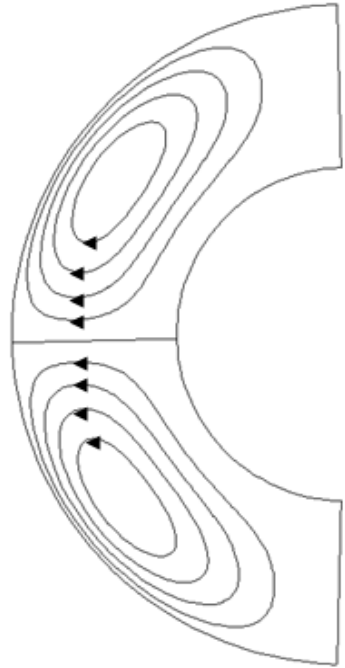

Fig. 4

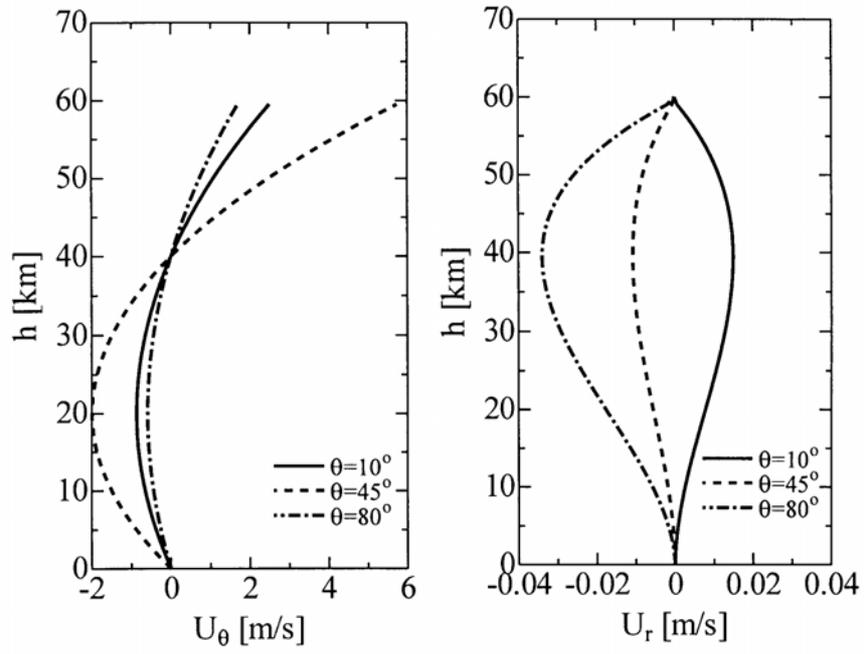

Fig. 5

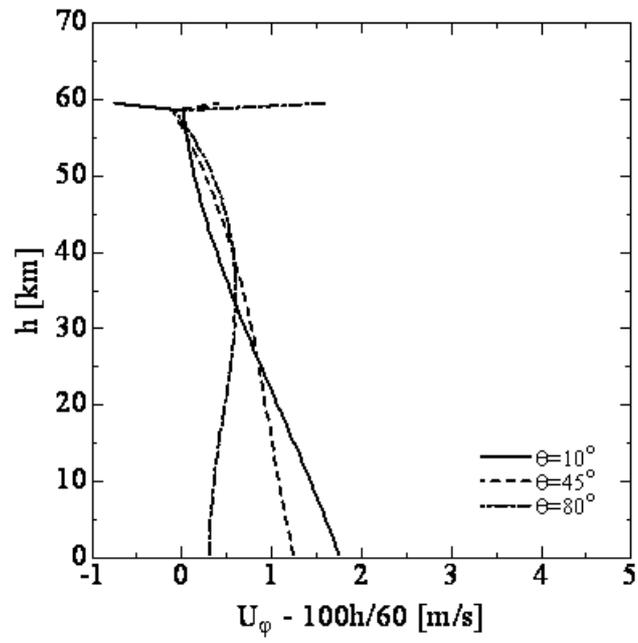

Fig. 6

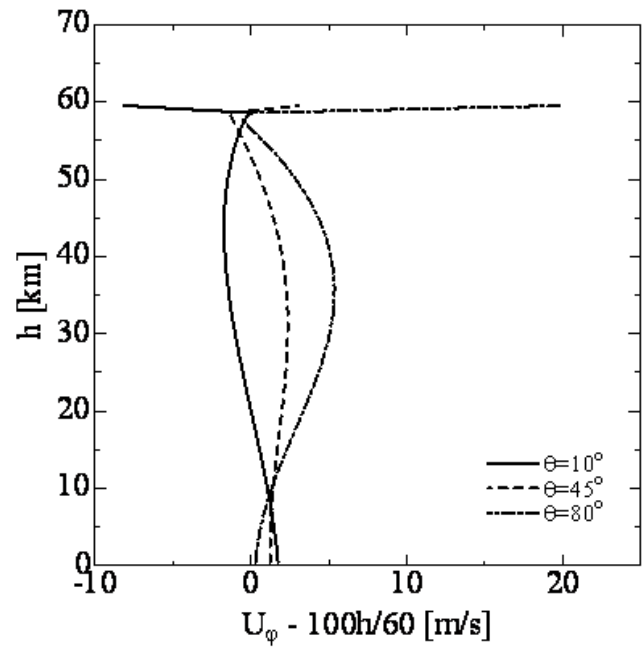

Fig. 7

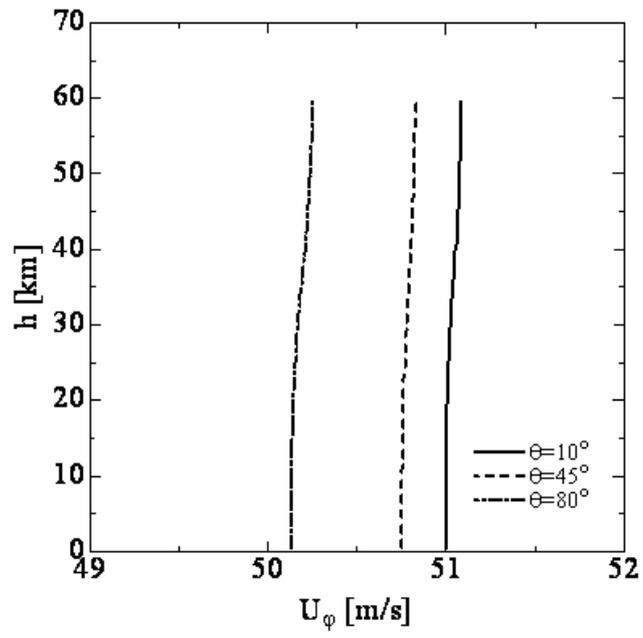

Fig. 8

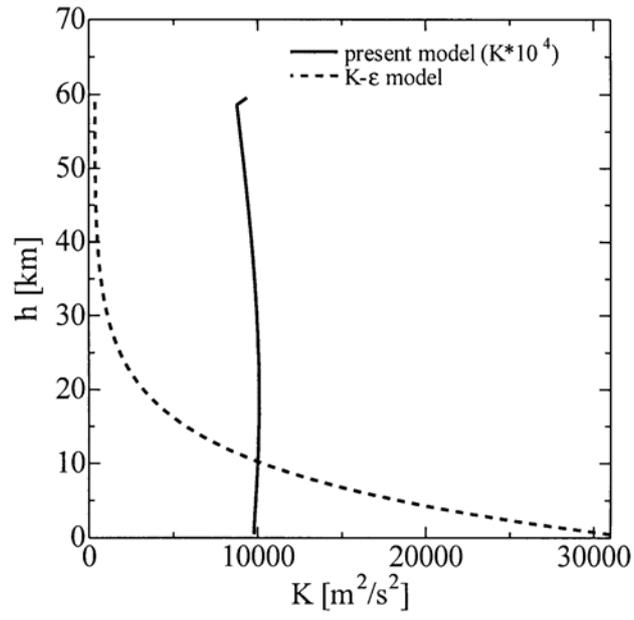

Fig. 9

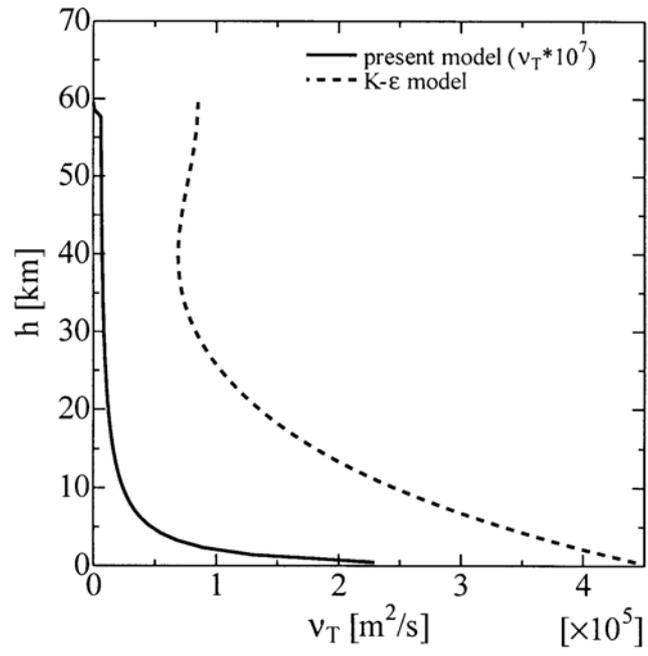

Fig. 10

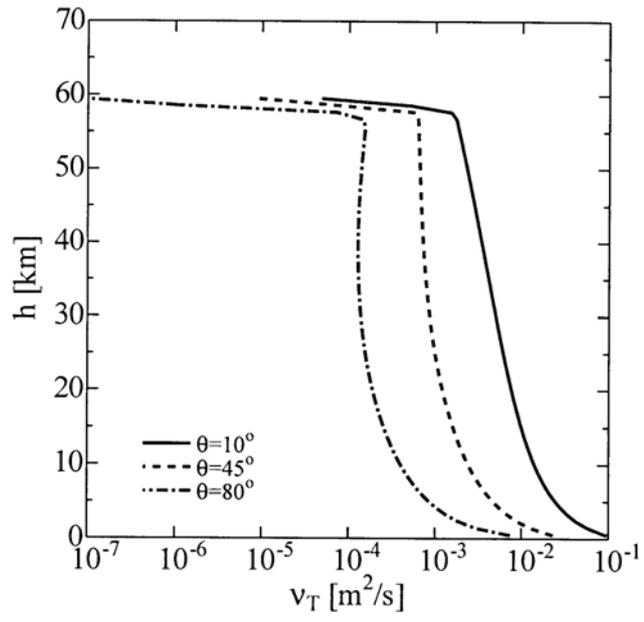

Fig. 11

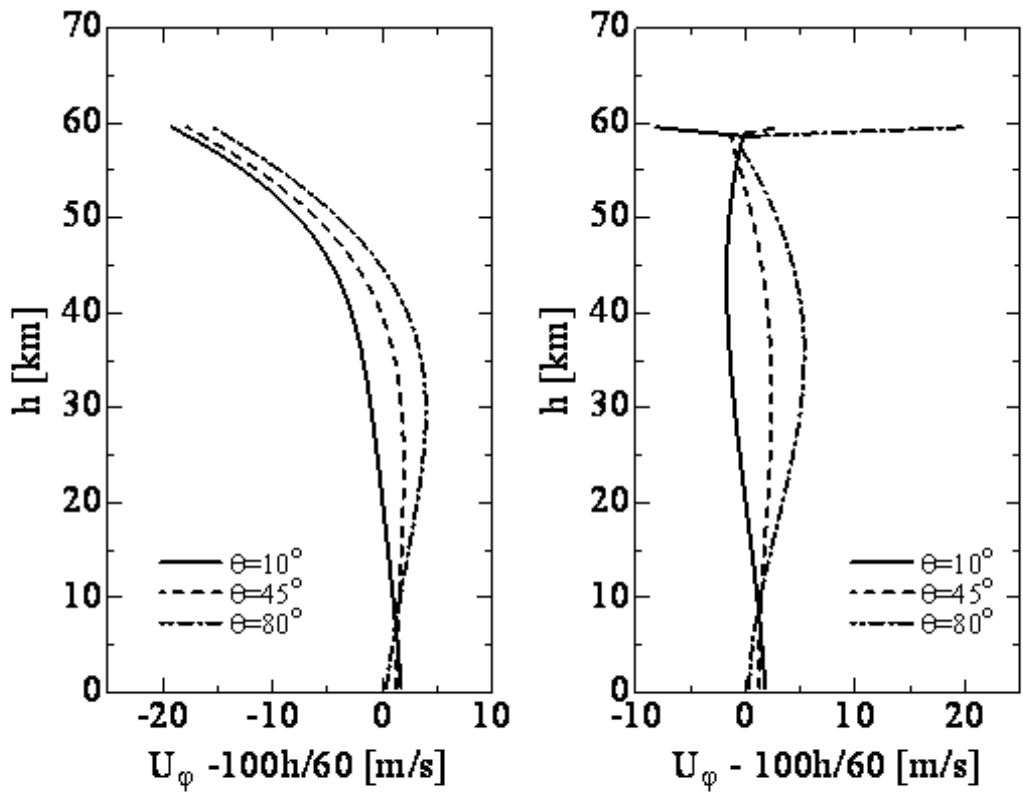

Fig. 12

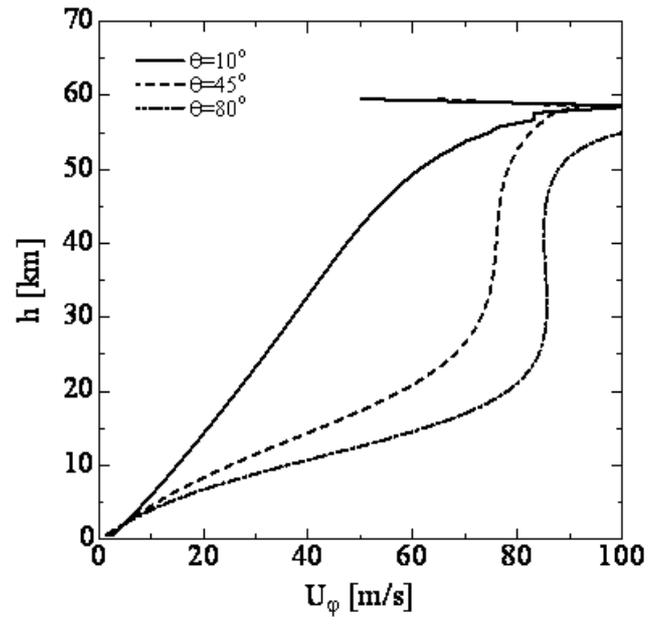

Fig. 13

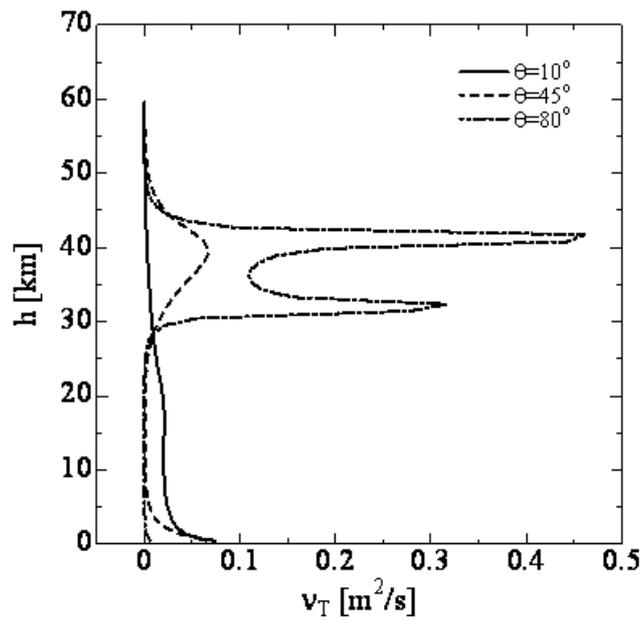